\documentclass{article}

\usepackage{amsmath, amssymb, amsthm}
\usepackage{graphicx}
\usepackage{hyperref}
\usepackage{cite}
\usepackage{geometry}
\usepackage{authblk}
\usepackage{booktabs} 
\usepackage{makecell} 
\usepackage{caption} 

\title{Comment on paper:

Position: Rethinking Post-Hoc Search-Based Neural Approaches for Solving Large-Scale Traveling Salesman Problems
}

\author{Yimeng Min}
\affil{Department of Computer Science \\
Cornell University \\
Ithaca, NY 14850 \\
Email: \href{mailto:min@cs.cornell.edu}{min@cs.cornell.edu}}

\date{}
\begin{document}

\maketitle

\begin{abstract}
We identify two major issues in the SoftDist paper: (1) the failure to run all steps of different baselines on the same hardware environment, and (2) the use of inconsistent time measurements when comparing to other baselines. These issues lead to flawed conclusions. When all steps are executed in the same hardware environment, the primary claim made in SoftDist is no longer supported.
\end{abstract}


\section{Introduction}
\label{sec:introduction}
In recent years, machine learning (ML) has emerged as a promising avenue for addressing optimization problems like the Travelling Salesman Problem (TSP).  ML techniques, particularly those involving neural networks and reinforcement learning, have shown potential in learning heuristics and patterns that can guide the search for optimal routes more efficiently. By leveraging data, ML models can improve the quality of solutions and reduce computation time. One notable approach is the use of heat map-based search, where ML models generate heat maps that highlight promising regions of the solution space. These heat maps are then used to focus the search process, potentially enhancing the efficiency and effectiveness of finding optimal or near-optimal solutions~\cite{utsp}.

Recently, the authors of paper~\cite{softdist} (referred to as SoftDist) discussed the neural approach and claimed: \begin{quote}
\textit{Our theoretical and experimental analysis raises doubts about the effectiveness of ML-based heat map generation. In support of this, we demonstrate that a simple baseline method can outperform complex ML approaches in heat map generation.}
\end{quote}

Here, however, \textbf{we show that the authors in SoftDist misconducted the experiments, leading to an unfair comparison and a flawed conclusion}. Using the correct comparison, the experimental results on large-scale instances like TSP-1000 contradict their claims. This seriously undermines the main assertion in their abstract.

In fact, their experiments demonstrate the power of unsupervised learning neural approaches on large instances like TSP-1000. Even after running a grid search on their proposed SoftDist model, their performance still falls short compared to unsupervised learning neural approaches.

\begin{figure}[h]
    \centering
    \includegraphics[width=0.9\textwidth]{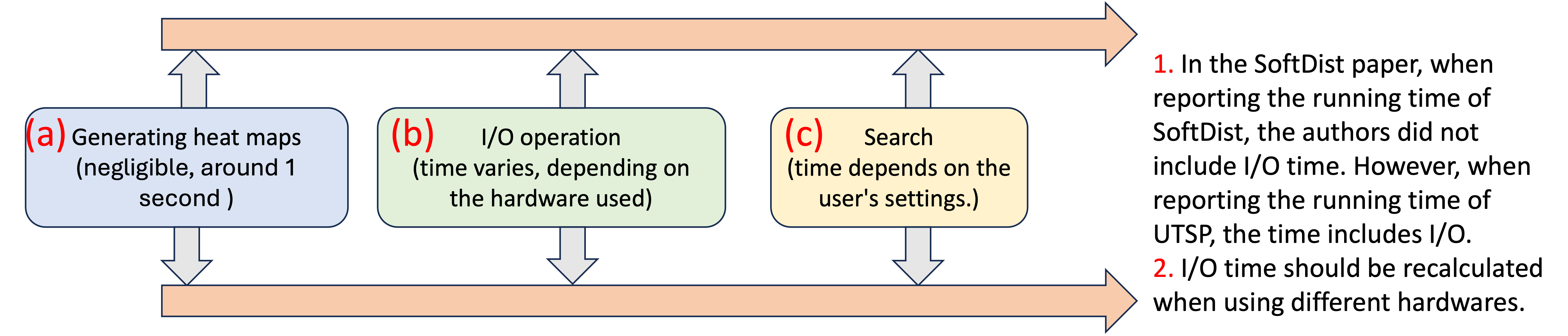} 
    \caption{The overall scheme of heat map-based search. In the SoftDist paper, when comparing the running time of different models, the authors calculate parts (a) and (c) for SoftDist, while the UTSP running time consists of parts (a), (b), and (c), leading to an unfair comparison.}
    \label{fig:fig1}
\end{figure}
\section{Background of Heat map-Based Search on TSP}
\label{sec:background}
Heat map-based search is a novel approach applied to the TSP, where heat maps are utilized to guide the search process. In this method, heat maps are generated to represent the solution space, highlighting areas of interest. These heat maps help guide the search, improving the efficiency and effectiveness of finding optimal or near-optimal solutions. This technique contrasts with traditional TSP methods by incorporating data-driven methods to direct the search, offering a unique way to tackle this well-known combinatorial optimization problem.

There are three steps in heat map-based search:
\begin{itemize}
    \item Generating the heat maps.
    \item Performing I/O operations to save or load the generated heat maps.
    \item Running the search.
\end{itemize}
The overall scheme of heat map-based search is shown in Figure~\ref{fig:fig1}. When comparing different methods, two factors are usually considered: the time required to build the heat maps and perform the search, and the quality of the TSP solutions. A method that provides better quality solutions within the same time cost is considered superior.

In the SoftDist paper, the authors compare the SoftDist method with other approaches using a search method proposed in UTSP~\cite{utsp}, as referenced in Table 3 of the SoftDist paper. Here, we show that there are two major problems in the SoftDist paper when comparing their proposed SoftDist method to the performance of UTSP~\cite{utsp}, a data-driven neural approach that leverages the power of unsupervised learning to generate heat maps for TSP without requiring labels.

\subsection{Problem 1}
In the SoftDist paper, the author states:
\begin{quote}
\textit{
    Our SoftDist heat map generation is performed on an NVIDIA A100 GPU. Due to its simplicity, the inference time is negligible (e.g., $<$ 0.1 seconds), and the GPU type has minimal impact. For fairness in comparison, all MCTS computations are conducted on an AMD EPYC 7V13 64-Core CPU @ 2.45GHz. We use 64 threads for TSP-500 and TSP-1000.
    }
\end{quote}
However, as shown in Figure~\ref{fig:fig1}, there are three steps in the heat map-based search. The authors only highlight running the search on the same AMD EPYC CPU, ignoring the other steps. Specifically, they rerun the search using their environment but do not account for the heat map generation part. In other words, the comparison is incomplete and biased, as the full computational process for heat map generation on the NVIDIA A100 GPU is not considered. This omission leads to misleading conclusions about the effectiveness and efficiency of the proposed method compared to other approaches.
\begin{table}[h]
    \centering
    \begin{tabular}{l|c|c|c}
        \toprule
        & \makecell{UTSP \\ (the measure in \\ UTSP paper)} & \makecell{SoftDist \\ (the measure in \\ SoftDist)} & \makecell{UTSP \\ (the measure in \\ SoftDist)} \\
        \midrule
        Generating heat maps & Included & Included & Included \\
        I/O operation & Included & Not Included & Not Included \\
        Search & Included & Included & Included \\
        Total time & 1s + 1.45m + I/O & 0s + 1.44m (w/o IO) & \textbf{1s + 1.45m (w/o IO)} \\
        Performance & 1.65\% & 1.73\% & \textbf{1.65\%} \\
        \bottomrule
    \end{tabular}
    \caption{Results on TSP-1000. The I/O operation always takes the same amount of time because we are saving or loading heat maps of the same size. Performance is measured by the gap with respect to the optimal solution; a lower performance value indicates a better solution. The performance values are directly taken from Table 3 in the SoftDist paper~\cite{softdist}. SoftDist underperforms compared to UTSP within nearly the same time budget.}
    \label{tab:comparison}
\end{table}
\subsection{Problem 2}
The time measurements used in SoftDist and UTSP are not the same, leading to an unfair comparison. In SoftDist, the total time considered by the authors includes only two parts: generating the heat map and running the search. However, when reporting UTSP’s total running time, the time includes three parts: generating the heat maps, I/O, and running the search. This discrepancy results in an unfair comparison between the methods.

To ensure a fair comparison, we recalculated the time cost using the SoftDist measurement. The results are shown in Table~\ref{tab:comparison}. When using the SoftDist measure, which only considers generating the heat maps and running the search, the UTSP method outperforms SoftDist with almost the same time budget. This suggests that the heat map generated using UTSP can better guide the search. Furthermore, since the generated heat maps should always have the same size, the I/O time for different heat map-based methods, when running in the same environment, should be the same.

Consequently, the main claim of the SoftDist paper, \begin{quote}\textit{we demonstrate that a simple baseline method can outperform complex ML approaches in heat map generation,}\end{quote} no longer holds.

It should also be highlighted that the authors of SoftDist conducted a grid search on their method while not performing any hyperparameter search on other methods. For example, they used a hyperparameter of 0.0066 for TSP-500 and 0.0051 for TSP-1000. This lack of consistent hyperparameter tuning across all methods further contributes to the unfair comparison and may skew the results in favor of SoftDist.

Overall, the performance on TSP-1000, using the same time measurement, shows that SoftDist~\cite{softdist}, even with hyperparameter grid search, still underperforms compared to data-driven neural approaches within almost the same time budget. Consequently, the claim in the SoftDist paper that their method can outperform ML approaches in heat map generation does not hold.

\section{Conclusion}
We identify two major issues in the SoftDist paper: (1) not running all steps of different baselines in the same hardware environment, and (2) using different time measurements when comparing to other baselines. These issues lead to flawed conclusions. When using the same hardware environment, the main claims of SoftDist are no longer supported.

These findings emphasize the importance of using comprehensive and fair evaluation metrics when comparing different methodologies. It also highlights the potential of unsupervised learning neural approaches in solving large-scale instances of TSP, demonstrating their superiority in both efficiency and effectiveness. Future research should focus on further refining these data-driven methods and exploring their applications in other complex combinatorial optimization problems.

\end{document}